\documentclass[12pt]{article}
\usepackage{amssymb}

\newcommand{\text}{\rm}
\newcommand{\be}{\begin{equation}}
\newcommand{\ee}{\end{equation}}
\newcommand{\bq}{\begin{eqnarray}}
\newcommand{\eq}{\end{eqnarray}}

\begin{document}

\begin{titlepage}
\title{{\bf  Aspects of Causality and  Unitarity and Comments on Vortex-like Configurations
in an Abelian Model with a Lorentz-Breaking Term}}
\author{{\bf A.P. Ba\^{e}ta Scarpelli$^{a}$}, {\bf H. Belich$^{a,b}$}, \\
{\bf J.L. Boldo$^{c}$} {\bf \thinspace and J.A. Helay\"el-Neto$^{a,b}\,$}
\thanks{{\tt E-mails:scarp@gft.ucp.br, belich@gft.ucp.br, jboldo@cce.ufes.br,
helayel@gft.ucp.br.}}\vspace{2mm} \\
{\small {\bf $^{a}$}Grupo de F\'{\i}sica Te\'{o}rica  Jos\'{e} Leite Lopes}\\
{\small Petr\'{o}polis, RJ, Brazil.}\vspace{2mm}\\
{\small {\bf $^{b}$}CBPF, Centro Brasileiro de Pesquisas F\'{\i }sicas }\\
{\small Rua Xavier Sigaud 150, 22290-180 Urca} \\
{\small Rio de Janeiro, RJ, Brazil.}\vspace{2mm}\\
{\small {\bf $^{c}$}UFES, Universidade Federal do Esp\'{\i }rito Santo}\\
{\small CCE, Departamento de F\'{\i}sica}\\
{\small Campus Universit\'{a}rio de Goiabeiras, 29060-900,}\\
{\small Vit\'{o}ria, ES, Brazil} \vspace{2mm}.}
\maketitle

\begin{abstract}
\noindent
The gauge-invariant Chern-Simons-type Lorentz- and CPT-breaking term is here 
reassessed and a spin-projector method is adopted to account for the breaking 
(vector) parameter. Issues like causality, unitarity, spontaneous gauge-symmetry 
breaking and vortex formation are investigated, and consistency conditions 
on the external vector are identified.
\end{abstract}
Pacs: 11.15.-g, 11.30.Cp, 11.15.Ex, 11.30.Er   
\end{titlepage}

\section{\ Introduction\-}

Symmetries are fundamental guides when one intends to systematize the study
of any theory. In this sense, Lorentz and CPT invariances acquire supreme
importance in modern Quantum Field Theory, both symmetries being respected
by the Standard Model for Particle Physics. A Standard Model description,
where possible violations of such invariances are considered, was developed
by Colladay and Kostelecky \cite{1}, \cite{2} and by Coleman and Glashow 
\cite{3}, \cite{4}. The main term that incorporates these features involves
the gauge field and has the Chern-Simons form 
\begin{equation}
\Sigma _{CS}=-\frac{1}{4}\int dx^{4}\epsilon ^{\mu \nu \alpha \beta }c_{\mu
}A_{\nu }F_{\alpha \beta },  \label{0.1}
\end{equation}
where $c_{\mu }$ is a constant four-vector that selects a space-time
direction \cite{5}-\cite{8}. One can easily show that such term originates a
vacuum optical activity. Astrophysical results \cite{9,10}, nevertheless,
contradict this possibility, putting very restrict limits on the magnitude
of the $c_{\mu }$ four-vector.

An interesting discussion was originated from the investigation on the
possibility that this Chern-Simons part be radiatively generated from the
fermionic sector of ordinary QED whenever an axial term, $b_{\mu }\bar{\Psi}%
\gamma ^{\mu }\gamma ^{5}\Psi $, that violates Lorentz and CPT symmetries,
is included \cite{11}-\cite{22}. The discussion took place around some
questions:

\begin{itemize}
\item  {Does this generated term depend on the regularization scheme?}

\item  {May the vanishing of this term be a result of gauge invariance and
unitarity requirements?}

\item  {Do the astrophysical observations impose limit on the radiative
correction generated by the axial term in the fermionic sector?}
\end{itemize}

As shown in \cite{19}, and argued in \cite{21}, the finite radiative
correction, $\Delta c_{\mu }$, comes from cancelation of divergences and,
therefore, is regularization-dependent. The condition for gauge invariance
can be stated in a weak way, since $b_{\mu }$ is a constant four-vector: it
is the action that must be invariant under this transformation and not
necessarily the Lagrangian density. It also means that it is not necessary
to be considered a source for the violating term. In \cite{21}, it was shown
that an indetermination in the radiative correction, $\Delta c_{\mu }$, is
not relevant for the physical content of the theory, since considering an
effective constant 
\begin{equation}
c_{\mu }^{eff}=(c+\Delta c+\delta c)_{\mu },  \label{0.2}
\end{equation}
where $\delta c_{\mu }$ is a finite counterterm (given some normalization
condition), one can always adjust the counterterm in order to obtain the
experimentally observed result.

We are then left with a careful analysis of limit situations, to which the
four-vector $c_{\mu }$ could be submitted, in order to verify if there is
physical consistency in some of these cases. In \cite{8}, the quantization
consistency of an Abelian theory with the inclusion of $\Sigma _{CS}$ is
thoroughly analyzed. The authors study the implications on the unitarity and
causality of the theory in cases where, for small magnitudes, $c_{\mu }$ is
time-like and space-like. The analysis shows that the behavior of these
Gauge Field Theories depends drastically on the space-time properties of $%
c_{\mu }$. According to \cite{8}, for a purely space-like $c_{\mu }$, one
finds a well-behaved Feynman propagator for the gauge field, and unitarity
and microcausality are maintained. On the other hand, a time-like $c_{\mu }$
spoils unitarity and causality.

In this work, we analyze the possibility of having consistency of the
quantization of an Abelian theory which incorporates the Lorentz- and
CPT-violating term of equation (\ref{0.1}), whenever gauge spontaneous
symmetry breaking (SSB) takes place. The analysis is carried out by pursuing
the investigation of unitarity and causality as read off from the
gauge-field propagators. We therefore propose a discussion at
tree-approximation, without going through the canonical quantization
procedure for field operators. In this investigation, we concentrate on the
analysis of the residue matrices at each pole of the propagators. Basically,
we check the positivity of the eigenvalues of the residue matrix associated
to a given simple pole in order that unitarity may be undertaken.
Higher-order poles unavoidably plague the theory with ghosts; this is why
our analysis of the residues is restricted to the case of the simple poles.
We shall find that only for $c_{\mu }$ space-like both causality and
unitarity can be ascertained. On the other hand, considering that SSB is
interesting in such a situation (since the mass generation mechanism induced
by the Higgs scalar presupposes that the theory is Lorentz invariant), we
obtain that, once Lorentz symmetry is violated, there is the possibility of
evading this mechanism, such that a gauge boson mass is not generated even
if SSB of the local U(1)-symmetry takes place.

In order to improve our comprehension of the physics presented by this
theory, we also study vortex-like configurations, analyzing the influence of
the direction selected by $c_{\mu }$ in space-time. The presence of the
Chern-Simons term produces interesting modifications on the equations of
motion that yield a vortex formation.

This work is outlined as follows: in Section 2, we study the SSB and present
our method to derive the gauge-field propagators. In Section 3, we set our
discussion on the poles and residues of the propagators. We study the
formation of vortices in Section 4, and, finally, in Section 5, we present
our Concluding Comments.

\section{The Gauge-Higgs Model}

We propose to carry out our analysis by starting off from the action 
\begin{equation}
\Sigma =\int d^{4}x\left\{ -\frac{1}{4}F_{\mu \nu }F^{\mu \nu }+\left(
D_{\mu }\varphi \right) ^{\ast }D^{\mu }\varphi -V\left( \varphi \right)
\right\} +\Sigma _{\psi }+\Sigma _{cs},  \label{1}
\end{equation}
where $\Sigma _{\psi }$ is some fermionic action (we do not introduce
fermions in our considerations here), 
\begin{equation}
\Sigma _{cs}=-\frac{\mu }{4}\int d^{4}x\varepsilon ^{\mu \nu \kappa \lambda
}v_{\mu }A_{\nu }F_{\kappa \lambda }  \label{2}
\end{equation}
is the Chern-Simons like term, $\mu $ is a mass parameter and $v_{\mu }$ is
an arbitrary four vector of unit length which selects a preferred direction
in the space-time ($c_{\mu }=\mu v_{\mu }$). The potential, $V$, given by 
\begin{equation}
V(\varphi )=m^{2}\left| \varphi \right| ^{2}+\lambda \left| \varphi \right|
^{4}
\end{equation}
is the most general Higgs-like potential in 4D. Setting suitably the
parameters such that the $\varphi $-field acquires a non-vanishing vacuum
expectation value (v.e.v), the mass spectrum of the photon would get shift
upon the spontaneous breaking of local gauge symmetry by means of such
v.e.v.. The Higgs field is minimally coupled to the electromagnetic by means
of its covariant derivative under U(1)-local gauge symmetry, namely 
\begin{equation}
D_{\mu }\varphi =\partial _{\mu }\varphi +ieQA_{\mu }\varphi .  \label{4}
\end{equation}

This symmetry is spontaneously broken, and the new vacuum is given by 
\begin{equation}
\langle 0|\varphi |0\rangle =a,
\end{equation}
where 
\begin{equation}
a=\left( -\frac{m^{2}}{2\lambda }\right) ^{1/2};\,\,\,\,m^{2}<0.
\end{equation}
As usual, we adopt the polar parametrization 
\begin{equation}
\varphi =\left( a+\frac{\sigma }{\sqrt{2}}\right) e^{i\rho /\sqrt{2}a},
\label{7}
\end{equation}
where $\sigma $ e $\rho $ are the scalar quantum fluctuations. Since we are
actually interested in the analysis of the excitation spectrum, we choose to
work in the unitary gauge, which is realized by setting $\rho=0$. Then, the
bilinear gauge action is given as below: 
\begin{equation}
\Sigma _{g}=\int d^{4}x\left\{ -\frac{1}{4}F_{\mu \nu }F^{\mu \nu }-\frac{%
\mu }{4}v_{\mu }A_{\nu }F_{\kappa \lambda }\varepsilon ^{\mu \nu \kappa
\lambda }+\frac{M^{2}}{2}A_{\mu }A^{\mu }\right\} ,  \label{9}
\end{equation}
where $M^{2}=2e^{2}Q^{2}a^{2}$.

Its is noteworthy to stress that the SSB introduces the mass term $M^{2}$ in
addition to the topological Lorentz-breaking term. $\mu $. As we shall see
throughout this Section, this term will simply shift the pole induced by $%
v^{\mu }$. If no SSB takes place, then $a=0$ and we reproduce the particle
spectrum given in \cite{8}. Another relevant issue to be tackled along this
Section regards the residues of the propagators at the poles, which inform
about the eventual existence of negative-norm 1-particle states. Later on,
suitable conditions on the parameters of the model will be adopted in order
that tachyon and ghost modes be suppressed from the spectrum.

We now rewrite the linearized action (\ref{9}) in a more convenient form,
namely 
\begin{equation}
\Sigma _{g}=\frac{1}{2}\int d^{4}xA^{\mu }\mathcal{O}_{\mu \nu }A^{\nu },
\label{10}
\end{equation}
where $\mathcal{O}_{\mu \nu }$ is the wave operator. The propagator is given
by 
\begin{equation}
\left\langle 0\right| T\left[ A_{\mu }\left( x\right) A_{\nu }\left(
y\right) \right] \left| 0\right\rangle =i\left( \mathcal{O}^{-1}\right)
_{\mu \nu }\delta ^{4}\left( x-y\right) .  \label{11}
\end{equation}
The wave operator can be written in terms of spin-projection operators as
follows: 
\begin{equation}
\mathcal{O}_{\mu \nu }=(\Box +M^{2})\theta _{\mu \nu }+M^{2}\omega _{\mu
\nu }+\mu S_{\mu \nu },  \label{12}
\end{equation}
where $\theta _{\mu \nu }$ and $\omega _{\mu \nu }$ are respectively the
transverse and longitudinal projector operators 
\begin{equation}
\theta _{\mu \nu }=g_{\mu \nu }-\frac{\partial _{\mu }\partial _{\nu }}{%
\Box }\;,\;\;\;\omega _{\mu \nu }=\frac{\partial _{\mu }\partial _{\nu }}{%
\Box },
\end{equation}
and 
\[
S^{\mu \nu }=\varepsilon ^{\mu \nu \kappa \lambda }v_{\kappa }\partial
_{\lambda }. 
\]
In order to invert the wave operator, one needs to add up other two new
operators, since the above ones do not form a closed algebra, as the
expression below indicates: 
\begin{eqnarray}
S_{\mu \alpha }S_{\,\,\,\,\,\nu }^{\alpha } &=&\left[ v^{2}\Box -\lambda
^{2}\right] \theta _{\mu \nu }-\lambda ^{2}\omega _{\mu \nu }-\Box
\Lambda _{\mu \nu }  \nonumber \\
&+&\lambda \left( \Sigma _{\mu \nu }+\Sigma _{\nu \mu }\right) \equiv f_{\mu
\nu },  \label{14}
\end{eqnarray}
with 
\begin{equation}
\Sigma _{\mu \nu }=v_{\mu }\partial _{\nu }\;,\;\;\lambda \equiv \Sigma
_{\mu }^{\;\mu }=v_{\mu }\partial ^{\mu }\;,\;\;\Lambda _{\mu \nu }=v_{\mu
}v_{\nu }.
\end{equation}
These results indicate that two new operators, namely, $\Sigma $ and $%
\Lambda $, must be included in order to have an operator algebra with closed
multiplicative rule. The operator algebra is displayed in Table 1. \vspace{%
2mm}

\begin{center}
\begin{tabular}{|c|c|c|c|c|c|c|}
\hline
& $\theta _{\,\,\,\,\,\nu }^{\alpha }$ & $\omega _{\,\,\,\,\,\nu }^{\alpha }$
& $S_{\,\,\,\,\,\nu }^{\alpha }$ & $\Lambda _{\,\,\,\,\,\nu }^{\alpha }$ & $%
\Sigma _{\,\,\,\,\,\nu }^{\alpha }$ & $\Sigma_\nu ^{\,\,\,\,\,\alpha}$ \\ 
\hline
$\theta _{\mu \alpha }$ & $\theta _{\mu \nu }$ & $0$ & $S_{\mu \nu }$ & $%
\Lambda _{\mu \nu }-\frac{\lambda }{\Box }\Sigma _{\nu \mu }$ & $\Sigma
_{\mu \nu }-\lambda \omega _{\mu \nu }$ & $0$ \\ \hline
$\omega _{\mu \alpha }$ & $0$ & $\omega _{\mu \nu }$ & $0$ & $\frac{\lambda 
}{\Box }\Sigma _{\nu \mu }$ & $\lambda \omega _{\mu \nu }$ & $\Sigma_{\nu
\mu}$ \\ \hline
$S_{\mu \alpha }$ & $S_{\mu \nu }$ & $0$ & $f_{\mu \nu }$ & $0$ & $0$ & $0$
\\ \hline
$\Lambda _{\mu \alpha }$ & $\Lambda _{\mu \nu }-\frac{\lambda }{\Box}%
\Sigma_{\mu \nu }$ & $\frac{\lambda }{\Box }\Sigma _{\mu \nu }$ & $0$ & $%
v^{2}\Lambda _{\mu \nu }$ & $v^{2}\Sigma _{\mu \nu }$ & $\lambda
\Lambda_{\mu \nu}$ \\ \hline
$\Sigma _{\mu \alpha }$ & $0$ & $\Sigma _{\mu \nu }$ & $0$ & $\lambda
\Lambda _{\mu \nu }$ & $\lambda \Sigma _{\mu \nu }$ & $\Lambda_{\mu \nu}
\Box$ \\ \hline
$\Sigma_{\alpha \mu}$ & $\Sigma_{\nu \mu} -\lambda \omega_{\mu \nu}$ & $%
\lambda \omega_{\mu \nu}$ & $0$ & $v^2 \Sigma_{\nu \mu}$ & $v^2 \Box
\omega_{\mu \nu}$ & $\lambda \Sigma_{\nu \mu}$ \\ \hline
\end{tabular}
\vspace{2mm}

Table 1: Multiplicative table fulfilled by $\theta ,\omega ,\,S,\,\Lambda $
and $\Sigma $. The products are supposed to obey the order ''row times
column''.
\end{center}

Using the spin-projector algebra displayed in Table 1, the propagator may be
obtained after a lengthy algebraic manipulation. Its explicit form in
momentum space is 
\begin{eqnarray}
\langle A_{\mu }A_{\nu }\rangle &=&\frac{i}{D}\left\{ -(k^{2}-M^{2})\theta
_{\mu \nu }+\left( \frac{D}{M^{2}}-\frac{\mu ^{2}(v\cdot k)^{2}}{%
(k^{2}-M^{2})}\right) \omega _{\mu \nu }\right.  \nonumber \\
&&\left. -i\mu S_{\mu \nu }-\frac{\mu ^{2}k^{2}}{(k^{2}-M^{2})}\Lambda _{\mu
\nu }+\frac{\mu ^{2}(v\cdot k)}{(k^{2}-M^{2})}\left( \Sigma _{\mu \nu
}+\Sigma _{\nu \mu }\right) \right\} ,  \label{13}
\end{eqnarray}
where $D(k)=(k^{2}-M^{2})^{2}+\mu ^{2}v^{2}k^{2}-\mu ^{2}(v\cdot k)^{2}.$

The result above enables us to set our discussion on the nature of the
excitations, read off as pole propagators, present in the spectrum. At first
sight, the denominator \ $(k^{2}-M^{2})$ appearing in connection with the
operators $\omega $, $\Lambda $, $\Sigma $, once multiplying the overall
denominator $D$, could be the origin for dangerous multiple poles that
plague the quantum spectrum with ghosts. For this reason, a careful study of
this question is worthwhile. With this purpose, it is advisable to split our
discussion into 3 cases: time-like, null (light-like) and space-like $v_{\mu
}$.

In the case $v_{\mu }$ is time-like, one can readily check that there will
always be possible to find momenta $k_{\mu }$ such that $\ k^{2}=M^{2}$
appears as a double pole in the transverse sector ($\theta $ and $S$) and a
triple pole in the $\omega $ -, $\Lambda $- and $\Sigma $- sectors. This
shows that in these situations non-physical states are present that
correspond to negative norm particle states. There is no need therefore to
discuss the residue matrix at these poles.

In the case $v_{\mu }$ is light-like, it can be seen that tachyonic poles (
that are simple poles ) always appear; this also invalidates the model in
this quantum version, for supraluminal excitations are always present in the
spectrum.

However, if $v_{\mu }$ is a space-like vector, no higher-order pole comes
out; it can be shown that $k^{2}=M^{2}$ is not a zero of $D(k)$. It only
appears as a simple pole for the $\omega $ -, $\Lambda $- and $\Sigma $-
sectors. So, the model exhibits non-tachyonic massive excitations associated
to 3 simple poles: 2 of them coming from $D(k)$, the other being $%
k^{2}=M^{2} $. The fact that only the space-like case is physically
acceptable confirms the detailed study carried out by Adam and Klinkhammer
in the work of ref. \ \cite{8}. Nevertheless, we should still investigate
the residue at these poles so as to be sure that no ghosts are present. This
shall carefully be done in the next section.

\section{Unitarity analysis in the space-like case}

Our present task consists in the check of the character of the poles present
for $v_{\mu }$ space-like. Knowing that 3 different poles show up, we have
to go through the study of the residue matrix of the vector propagator at
each of its (time-like) poles $k^{2}=M^{2}$, $k^{2}=\tilde{m}_{1}^{2}$ and $%
k^{2}=\tilde{m}_{2}^{2}$, where $\tilde{m}_{1}$ and $\tilde{m}_{2}$
correspond to the zeroes of $D(k)$, that is, $M^{2}$, $\tilde{m}_{1}^{2}$
and $\tilde{m}_{2}^{2}$ are the physical masses at the tree approximation.

To infer about the physical nature of the simple poles, we have to calculate
the eigenvalues of the residue matrix for each of these poles. This is done
in the sequel.\ Before quoting our results, we should say that, without loss
of generality, we fix our external space-like\ vector as given by $v^{\mu
}=(0;0,0,1)$. The momentum propagator, $k^{\mu }$, is actually a
Fourier-integration variable, so we are allowed to pick a representative
momentum whenever $k^{2}>0$. We pursue our analysis of the residues by
taking $k^{\mu }=(k^{0};0,0,k^{3})$.

With $k_{0}^{2}=m_{1}^{2}$, we have that 
\begin{equation}
m_{1}^{2}=\frac{2\left( M^{2}+k_{3}^{2}\right) +\mu ^{2}+\mu \sqrt{\mu
^{2}+4\left( M^{2}+k_{3}^{2}\right) }}{2}\mathrm{;}
\end{equation}
the residue matrix reads as below 
\begin{equation}
R_{1}=\frac{1}{\sqrt{\mu ^{2}+4\left( M^{2}+k_{3}^{2}\right) }}\left( 
\begin{array}{cccc}
0 & 0 & 0 & 0 \\ 
0 & m_{1}^{2}-\left( M^{2}+k_{3}^{2}\right) & i\mu m_{1} & 0 \\ 
0 & -i\mu m_{1} & m_{1}^{2}-\left( M^{2}+k_{3}^{2}\right) & 0 \\ 
0 & 0 & 0 & 0
\end{array}
\right) .
\end{equation}
We calculate its eigenvalues and find only a single non-vanish eigenvalue: 
\begin{equation}
\lambda =\frac{2\left| m_{1}\right| }{\sqrt{\mu ^{2}+4\left(
M^{2}+k_{3}^{2}\right) }}>0
\end{equation}

The same procedure and the same conclusions hold through for the second zero
of $D(k)$ $\left( k^{2}=\tilde{m}_{2}^{2}\mathrm{{\ with }k_{0}^{2}=m_{2}^{2}%
}\right) $: 
\begin{equation}
k_{0}^{2}=m_{2}^{2}=\frac{2\left( M^{2}+k_{3}^{2}\right) +\mu ^{2}-\mu \sqrt{%
\mu ^{2}+4M^{2}}}{2}\mathrm{;}
\end{equation}

there comes out a unique non-vanishing eigenvalue $\left( \lambda =\frac{%
2\left| m_{2}\right| }{\sqrt{\mu ^{2}+4\left( M^{2}+k_{3}^{2}\right) }}%
>0\right) $ as above.

The calculations above confirm the results found by the autors of ref.\cite
{8}: for a space-like $v^{\mu }$, the pole of $D(k)$ respect causality (they
are not tachyonic) and correspond to physically acceptable $1-$particle
states with $1$ degree of freedom, since the residue matrix exhibits a
single positive eigenvalue

Finally, we are left with the consideration of the pole $k_{0}^{2}=\left(
M^{2}+k_{3}^{2}\right) $. The residue matrix reads as follows: 
\begin{equation}
R_{M}=\left( 
\begin{array}{cccc}
-\frac{\mu ^{2}}{M^{2}}k_{3}^{2}\left( M^{2}+k_{3}^{2}\right)  & 0 & 0 & -%
\frac{\mu ^{2}}{M^{2}}\left| k_{3}\right| \left( M^{2}+k_{3}^{2}\right) ^{%
\frac{3}{2}} \\ 
0 & 0 & 0 & 0 \\ 
0 & 0 & 0 & 0 \\ 
-\frac{\mu ^{2}}{M^{2}}\left| k_{3}\right| \left( M^{2}+k_{3}^{2}\right) ^{%
\frac{3}{2}} & 0 & 0 & -\frac{\mu ^{2}}{M^{2}}\left( M^{2}+k_{3}^{2}\right)
^{2}
\end{array}
\right) ,
\end{equation}
and again we have obtained only a non-vanish eigenvalue: $\lambda =\frac{1}{%
M^{2}}\left( M^{2}+2k_{3}^{2}\right) >0$. This opens up a very interesting
conclusion: the $M^{2}$-pole, appearing in the longitudinal sector $\left(
w_{\mu \nu }\right) $, describes a physically realizable scalar mode. We are
before a very peculiar result: The vector potential accommodates $3$
physical excitations ( with masses $m_{1}^{2}$, $m_{2}^{2}$, and $M^{2}$),
each of them carrying a single degree of freedom; so, the external
background influences the gauge field by drastically changing its physical
content: instead of describing a $3-$degree of freedom massive excitation,
it rather describes \ $3$ different massive excitations, each carrying one
physical degree of freedom.

We would like to report on one more possibility. As we know, the Higgs
mechanism for mass generation for gauge bosons presupposes Lorentz
invariance of the theory. This is no longer our case. So, we want to exhibit
that, for a fixed background space-like vector, $v^{\mu },$ there may appear
massless modes depending on the direction of the wave propagation. Indeed,
the condition for a massless pole, $D(k)=0$ with $\ k^{2}=0$, can be written
as

\bigskip\ 
\begin{equation}
c\cdot k=\pm M^{2}.  \label{34}
\end{equation}

\bigskip Taking a space-like $c_{\mu }$ of the form $c_{\mu }=(0;\vec{c})$,
the condition above reads

\[
\vec{c}\cdot \vec{k}=\mp M^{2}. 
\]

With $K^{2}=0$, $|\vec{k}|=k^{0}$, whenever $\ k^{0}$ $\rangle $ $0$; then,
we see that

\[
\vec{c}\cdot \hat{k}=-\frac{M^{2}}{k^{0}}. 
\]

So, given $\vec{c}$, we can always find a\ $k^{\mu }$ such that $k^{2}=0$ is
compatible with the condition above; for this to take place, the propagation
must be along a direction with an angle bigger than $90^{o}$. The conclusion
is that, according to the direction of the wave propagation, a massless pole
shall always show up. This confirms the breaking of isotropy and illustrates
that, despite spontaneous breaking of a local symmetry, massless excitations
may be present in the spectrum.

After the technical details exposed previously, we should clarify better our
analysis of the unitarity. In the paper of ref. \cite{8}, the authors raise
the question of the unitarity and they conclude that, exclusively for a
space-like $v^{\mu }$, the Hamiltonian admits a semi-positive self-adjoint
extension, giving therefore rise to a unitary time evolution operator.

Here, the unitarity alluded to is not in the sense of a self-adjoint
extension, but rather in the framework of the Hilbert space of particle
states. Our analysis reveals the existence of 1-particle states with
negative norm square, i.e., 1-particle ghost states, whenever $v^{\mu }$ is
time- or light-like. On the other hand, when $v^{\mu }$ is space-like, the
poles of the vector propagator are physically acceptable and the model may
be adopted as a consistent theory.

\section{A discussion on vortex-like configurations}

Once our discussion on the consistency of the quantum-mechanical properties
of the model has been settled down, we would like to address to an issue of
a classical orientation, namely, the reassessment of vortex-like
configurations in the presence of Lorentz-breaking term as the one we tackle
here.

In our case, with the Chern-Simons-like term included, we get, from the
action (3), the equations of motion 
\begin{equation}
D^{\mu }D_{\mu }\varphi =-m^{2}\varphi -2\lambda \varphi |\varphi |^{2}
\label{eu}
\end{equation}
and 
\begin{equation}
ie(\varphi \partial ^{\mu }\varphi ^{\ast }-\varphi ^{\ast }\partial ^{\mu
}\varphi )+2e^{2}A^{\mu }|\varphi |^{2}+\mu \varepsilon ^{\mu \nu \kappa
\lambda }v_{\nu }\partial _{\kappa }A_{\lambda }=\partial _{\nu }F^{\mu \nu
},  \label{euu}
\end{equation}
so that we can explicitly derive the modified Maxwell equations 
\begin{equation}
\mathbf{\nabla .E}=-ie(\varphi \dot{\varphi}^{\ast }-\varphi ^{\ast }\dot{%
\varphi})+2e^{2}|\varphi |^{2}\Phi -\mu \mathbf{v\cdot B}  \label{divE}
\end{equation}

\bigskip 
\begin{equation}
\mathbf{\nabla }\times \mathbf{E=}-\frac{\partial \mathbf{B}}{\partial t}
\label{rotE}
\end{equation}

and

\begin{equation}
\mathbf{\nabla .B=}0  \label{mon}
\end{equation}
\begin{eqnarray}
-\frac{\partial \mathbf{E}}{\partial t}+\mathbf{\nabla }\times \mathbf{B}
&=&ie(\varphi \mathbf{\nabla }\varphi ^{\ast }-\varphi ^{\ast }\mathbf{%
\nabla }\varphi )-2e^{2}|\varphi |^{2}\mathbf{A}  \nonumber \\
&-&\mu v_{0}\mathbf{B}+\mu \mathbf{v}\times \mathbf{E}.  \label{37}
\end{eqnarray}

Before going on to analyses vortex configurations, we would like to handle
the modified Maxwell equations above (eqs. (\ref{divE})-(\ref{37})) to
understand that there is no room for a magnetic monopole once the
Lorentz-breaking Chern-Simons term is switched on. For this purpose, we
remove the charged scalar field and see that the presence of a static
monopole immediately leads to

\begin{equation}
v_{0}B=v\times E  \label{37a}
\end{equation}

Now, by applying the operator\textbf{\ }$\mathbf{\nabla }\cdot $ to this
equation, we come to a direct contradiction with eq. (\ref{mon}). So, the
modified Maxwell equations (\ref{divE})-(\ref{37}) do not support the
presence of a Dirac-like magnetic monopole.

To analyse the vortex-type solutions, we consider a scalar field in $2$%
-dimensional space. The asymptotic solution that is proposed to be a circle $%
(S^{1})$ 
\begin{equation}
\varphi =ae^{in\theta };~~~~~~(r\rightarrow \infty ),  \label{38}
\end{equation}
where $r$ and $\theta $ are polar coordinates in the plane, $a$ is a
constant and $n$ is an integer. The gauge field assumes the form 
\begin{equation}
\mathbf{A}=\frac{1}{e}\mathbf{\nabla }(n\theta );~~~~~~(r\rightarrow \infty
),  \label{39}
\end{equation}
or, in term of its components: 
\begin{equation}
A_{r}\rightarrow 0,~~~A_{\theta }\rightarrow -\frac{n}{er}%
;~~~~~~(r\rightarrow \infty ).  \label{40}
\end{equation}

will be analyzed with our solution of the field $\Phi .$

The breaking of Lorentz covariance prevents us from setting $A_{\mu }$ as a
pure gauge at infinity, as usually done for the Nilsen-Olesen vortex. This
means that $A^{0}=\Phi (r)$, as $r\rightarrow \infty $. The asymptotic
behavior of $\Phi $ shall be fixed by the field equations, as shown in the
sequel. Returning to our problem, in this situation, the magnetic field
presents a cylindrical symmetry and 
\begin{equation}
\varphi =\chi (r)e^{in\theta }.
\end{equation}
To avoid singularity for $r\rightarrow 0$ and to keep an asymptotic
solution, we make 
\begin{equation}
lim_{r\rightarrow 0}\chi (r)=0
\end{equation}
and 
\begin{equation}
lim_{r\rightarrow \infty }\chi (r)=a.
\end{equation}
In the static case, equation (\ref{eu}), after summing over the components,
becomes 
\begin{equation}
\frac{1}{r}\frac{d}{dr}\left( r\frac{d\chi }{dr}\right) -\left[ \left( \frac{%
n}{r}+eA\right) ^{2}+m^{2}+2\lambda \chi ^{2}-e^{2}\Phi ^{2}\right] \chi =0,
\label{ce}
\end{equation}
while the modified Maxwell equations take the form 
\begin{equation}
{\nabla }^{2}\Phi +2e^{2}\chi ^{2}\Phi -\mu \mathbf{v\cdot B}=0  \label{m1}
\end{equation}
and 
\begin{equation}
\frac{d}{dr}\left( \frac{1}{r}\frac{d}{dr}\left( rA\right) \right) +2e\chi
^{2}\left( \frac{n}{r}-eA\right) -\mu v_{3}\frac{d\Phi }{dr}=0.  \label{m2}
\end{equation}
In the asymptotic region, equations (\ref{m1}) and (\ref{m2}) become 
\begin{equation}
{\nabla }^{2}\Phi -2a^{2}e^{2}\Phi =0
\end{equation}
and 
\begin{equation}
\frac{d}{dr}\left( \frac{1}{r}\frac{d}{dr}\left( rA\right) \right)
-2e^{2}a^{2}A-\mu v_{3}\frac{d\Phi }{dr}=0,
\end{equation}
where $\mathbf{B}$ has been set to zero, for $\mathbf{A}$ is a gradient at
infinity. We then find 
\begin{equation}
\Phi =Ce^{-\sqrt{2a^{2}e^{2}}~r}  \label{45a}
\end{equation}
and 
\begin{eqnarray}
A(r) &=&CK_{1}\left( \sqrt{2}a\left| e\right| r\right) +  \nonumber \\
&-&i\,\sqrt{2}\mu v_{3}\,aeK_{1}\left( \sqrt{2}a\left| e\right| r\right)
\int {rdrI}_{1}\left( \sqrt{2}a\left| e\right| r\right) {e^{-\sqrt{2}a\left|
e\right| r}}.
\end{eqnarray}
So, both $\Phi $ and $A$\ falls down to zero exponentially in the asymptotic
region. Note that, asymptotically the complex scalar field $\varphi =\chi
(r)e^{in\theta }$ \ goes to a \ non-trivial vacuum and becomes $\varphi
=ae^{in\theta }$ . Then the topology of the vacuum manifold is $S^{1}$.

A relevant discussion at this point is the issue of the stability of the
vortex configuration we have identified. This question has to be answered if
we have some elements about the energy of the system. Following the results
of the work of (\cite{9}), we understand that, once $v^{\mu }$ is chosen to
be space-like (and, according to the results of our discussion in Sec.3,
this is the unique sensible situation), the energy is limited from below,
which assigns to our vortex the status of stable configuration. \ 

More generally than in the case of Nielsen-Olesen vortices \cite{23}, eq. (%
\ref{m1}) plays an important role as long as the electric field is
concerned. If the magnetic field vortex (supposed such that $\mathbf{B}=B%
\mathbf{\hat{z}}$) is orthogonal to the external vector $\mathbf{v}$, then $%
\Phi =0$ is always a trivial solution that is compatible with the whole set
of field equations.

However, whenever $\mathbf{v\cdot B}\neq 0$, $\Phi $ must necessarily be
non-trivial, and an electric field appears along with the magnetic flux. If
this is the situation, in the asymptotic region $\Phi $ falls off
exponentially, as exhibited in eq. (\ref{45a}).

The appearance of an electrostatic field attached to the magnetic vortex,
whenever $\mathbf{v\cdot B}\neq 0$, is not surprising. Its origin may be
traced back to the Lorentz-breaking term: indeed, being a Chern-Simons-like
term, the electrostatic problem induces a magnetic field and the
magnetostatic regime demands an electric field too. So, a non-vanishing $%
\Phi $, therefore a non-trivial $\mathbf{E}$ response to the Chern-Simons
Lorentz-breaking term.

\section{Concluding Comments}

The main purpose of our work is the investigation of two aspects: the first
one is the quantization consistency of an Abelian model with violation of
Lorentz and CPT symmetries, contemporarily with the spontaneous breaking of
gauge symmetry. The other one concerns the study of classical vortex-like
configurations eventually present in such a model.

The analysis carried out with the help of the propagators, derived thanks to
an algebra of extended spin operators, reveals that unitarity is always
violated for $v^{\mu }$ time-like and null. Whenever the external vector is
space-like, physically consistent excitations can be found that present a
single degree of freedom each.

The analysis of the classical vortex-like configurations shows some
interesting aspects. First, if the magnetic field vortex is orthogonal to
the plane which contains the constant vector $v^{\mu }$, then a trivial
solution for the scalar potential, $\Phi =0$, is allowed. In this case, the
vortex configuration will be similar to the one of the usual Abelian model.
However, if $\mathbf{v}\cdot \mathbf{B}\neq 0$, we have a non-trivial
solution for $\Phi $ and an electric field appears in connection with the
magnetic flux. As we have already pointed out, the appearance of an electric
field attached to the magnetic vortex is not surprising. It is the
counterpart of what happens in a Chern-Simons theory in three dimensions,
where the electrostatic problem induces a magnetic field and the
magnetostatic regime demands an electric field too.

In connection with this phenomenon, the analysis of the dynamics of
electrically charged particles, magnetic monopoles and neutrinos in the
region outside the vortex core becomes a well-motivated idea, for the
presence of the electric field interferes now (at least for charged
particles and monopoles) and alter our knowledge about the concentration of
the particles in the region dominated by the vortex.

Finally, in view of the interesting results presented by Berger and
Kostelek\'{y} in the paper of ref. \cite{24}, it would be a relevant task to
incorporate the (gauge-invariant) Lorentz-breaking term in the action (\ref
{0.1}), in a supersymmetric framework and therefore to study the gaugino
counterpart of the action term given by equation (\ref{2}). Results in this
direction shall soon be presented elsewhere \cite{25}.

\begin{center}
\vspace{5mm}
\end{center}

{\Large Acknowledgments}

The authors are grateful to M. M. Ferreira Junior and O. Piguet for very
clarifying and detailed discussions. They also express their gratitude to
CNPq for the invaluable financial help

\begin{center}
\vspace{5mm}
\end{center}

\end{document}